\documentclass[preprint2]{aastex}

\shorttitle{Low-mass visual companions to nearby G-dwarfs}
\shortauthors{Tokovinin}

\begin{document}

\title{Low-mass visual companions to nearby G-dwarfs}

\author{Andrei Tokovinin}
\affil{Cerro Tololo Inter-American Observatory, Casilla 603, La Serena, Chile}
\email{atokovinin@ctio.noao.edu}

\begin{abstract}
Complete census of wide visual  companions to nearby G-dwarf stars can
be  achieved  by  selecting  candidates from  the  2MASS  Point-Source
Catalog and checking their  status by second-epoch imaging.  Such data
are obtained for  124 candidates with separations up  to $20''$, 47 of
which are  shown to  be new physical  low-mass stellar  companions.  A
list  of  visual  binaries  with  G-dwarf  primaries  is  produced  by
combining  newly  found  companions  with  historical  data.   Maximum
likelihood  analysis leads  to the  companion frequency  of  $0.13 \pm
0.015$  per decade  of separation.   The mass  ratio  is distributed
almost uniformly,  with a power-law  index between $-0.4$ and  0.  The
remaining  uncertainty in  the index  is  related to  modeling of  the
companion detection threshold in  2MASS.  These findings are confirmed
by  alternative analysis of  wider companions  in 2MASS,  removing the
contamination  by background stars  statistically.  Extension  of this
work  will lead  to a  complete detection  of visual  companions  -- a
necessary step towards  reaching unbiased multiplicity statistics over
the full  range of orbital periods and,  eventually, understanding the
origin of multiple systems.
\end{abstract}

\keywords{stars: binaries}

%-------------------------------------------------------------
\section{Introduction}
\label{sec:intro}

Attention  to  low-mass stars  in  the  solar  neighborhood is  mostly
inspired by  searches for exo-planets. Looking  for stellar companions
is  less  popular  nowadays,  although  solid empirical  data  on  the
distribution of  periods, mass ratios, and hierarchies  are still much
needed for  understanding the origin  of stars, multiple  systems, and
planets   \citep{Bate08}.   Even   within  the   25-pc   distance  the
multiplicity statistics of G-dwarfs is still being updated and revised
\citep{Raghavan10}.  An order-of-magnitude  larger sample is required,
however,  to study  detailed distributions  of orbital  parameters and
higher-order hierarchies (triples, quadruples, etc.).

Periods of  binary systems range  from half-day to millions  of years.
Discovery of  all companions over the entire  parameter space requires
combination  of complementary  techniques.  Precise  radial velocities
cover periods up to few years, but information at longer periods is still
dominated  by visual discoveries  over the  last two  centuries.  These
data,  collected   in  the  Washington  Double   Star  Catalog,  WDS
\citep{WDS},  are  notoriously  incomplete,  especially  for  low-mass
companions.

The Two  Micron All-Sky Survey,  2MASS \citep{2MASS} covers  the whole
sky in  the $JHK_s$ near-IR  bands. Its sensitivity is  sufficient for
detecting  even  a  0.08-$M_\odot$  companion  to a  G-dwarf  star  at
60\,pc. The problem is  in separating true ({\it physical}) companions
from  the  background  stars  ({\it optical}  companions).   At  large
separations,  physical companions  can be  identified by  their common
proper motion  (PM) \citep{Lepine07,Makarov08}.  However,  the PM data
are mostly based on archival photographic images where the vicinity of
bright  primary  stars  is  contaminated  by  their  halo,  preventing
detection of  very faint companions  (the magnitude difference  in the
visible  is larger than  in the  IR).  Meanwhile,  physical companions
should  dominate  over background  interlopers  at small  separations.
IR imaging  is used to find visual  companions to exoplanet
host stars  by, e.g. \citet{Mugrauer09}.   \citet{Kirkpatrick10} took
second-epoch  images   of    10\%   of  sky  and  found   many  new
common-proper-motion companions by comparison with the 2MASS.

The purpose  of this  work is  to explore the  potential of  2MASS for
reaching a complete census of stellar companions to nearby dwarfs with
separations  from a  few  arcseconds  to $20''  -  30''$.  At  smaller
separations,                  adaptive-optics                  imaging
\citep{ST02,Kraus08,MH09,Chauvin10} can be used, at larger separations
the existing PM data may be sufficient.  I take second-epoch images of
carefully  selected companion candidates  from the  2MASS Point-Source
Catalog  (hereafter  PSC)  \citep{2MASS}  to  determine  their  status
(physical or optical). The detection limit of the PSC is modeled.  The
data set on visual companions  in the selected range of separations is
complemented by the  new discoveries from 2MASS and  used to study the
distribution of the companion mass  ratio, $f(q)$.  I show that $f(q)$
is approximately uniform.  This  refers to wide binaries studied here;
a  typical $10''$  binary at  60\,pc distance  has semi-major  axis of
600\,AU and an orbital period on the order of 15\,000~yr.

New companions  to G-dwarfs  found here extend  their number  by 55\%,
complementing the census in the  low-mass regime. About 1/3 of the new
binaries are  higher-order multiples.  More  companion candidates from
the PSC wait to be confirmed by contemporary imaging.  This study thus
contributes  to  a  larger  task of  obtaining  unbiased  multiplicity
statistics of nearby G-dwarfs. 

The  sample of nearby  G-dwarfs and  criteria for  selecting candidate
companions      are      presented      in      Section~\ref{sec:samp}.
Section~\ref{sec:data} describes the observations and their processing.
The results  are presented  in Section~\ref{sec:res}, followed  by the
statistical  analysis in Section~\ref{sec:stat}.  The paper  closes by
discussion in Section~\ref{sec:disc}.

%-------------------------------------------------------------
\section{The sample}
\label{sec:samp}

The targets chosen  for this survey belong to a  large sample of nearby
solar-type dwarfs ({\it Nsample})  selected from the 
{\it   Hipparcos}   catalog   \citep{HIP}   in  its   latest   version
\citep{vLHIP} by the following criteria.
\begin{enumerate}
\item
Trigonometric parallax $\pi_{\rm HIP}  < 15$\,mas (within 67\,pc from
the Sun, distance modulus  $<4.12^m$). 

\item
Color $0.5  < V-I <  0.8$ (this corresponds approximately  to spectral
types from F5V to K0V).

\item
Unevolved, satisfying the condition  $M_{\rm Hp} > 9(V-I) - 3.5$,
where $M_{\rm Hp}$ is the absolute magnitude in the {\it Hipparcos} band
calculated with  $\pi_{\rm HIP}$. 

\end{enumerate}
There  are  5040 catalog  entries  satisfying  these conditions.   The
selection criteria may introduce  some bias with respect to unresolved
binaries, to be  addressed in the final statistics  but irrelevant for
the  present study.  Systems containing  white dwarfs  should be
removed  from  the  statistics,  as  the present-day  G-dwarfs  are  not
original primaries.   Removal of the secondary  components of binaries
with separate HIP numbers and 6 stars with parallax errors larger than
7.5\,mas leaves 4915 primary  targets. This sample includes many stars
monitored for  exoplanets and young  stars in the  solar neighbourhood
(X-ray sources, members of T-associacions).

All point sources within 2.5$'$  radius from each target are extracted
from the  PSC.  The  number of companions  around each  target, $N^*$,
varies strongly and  is a good measure of  crowding.  The total number
of stars around primaries is 114404, or $N^* = 23.3$ on average.  Most
of these companions are faint.

Trying  to isolate  potential  physical companions,  I consider  only
point sources within  $20''$ from primary targets which  do not belong
to crowded fields, $N^* < 100$.  Primary components not found in 2MASS
(within $6''$) are skipped.  This leaves 5347 stars, of which 4281 are
primaries, 1066 are candidate secondaries.  Among the secondaries, 238
are known companions listed in the WDS \citep{WDS}.

\begin{figure}[ht]
\epsscale{1.0}
\plotone{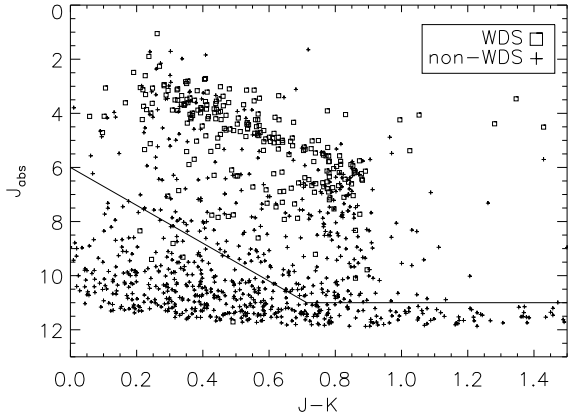}
\caption{The  $(J_{\rm abs},  J-K)$  CMD of  PSC candidate  companions
  within $20''$ from the {\it Nsample} targets.
 \label{fig:cmd}}
\end{figure}

Observations were  conducted to determine which of  the companions are
physical. New  imaging covers the  right ascension zone from  $2^h$ to
$18^h$ and  declinations south of  $+20^\circ$, 0.447 fraction  of the
whole sky.  As the majority  of the candidate companions are too faint
to be  on the Main-Sequence (MS), an additional  selection criterion is
imposed,
\begin{equation}
J_{\rm abs} < 6 + 7(J-K) \;\;\; {\rm and}  \;\;\; J_{\rm abs} < 11 .
\label{eq:crit}
\end{equation}

Thus,  366  potential companions  are  above  the  solid line  on  the
$(J_{\rm    abs},    J-K)$    color-magnitude   diagram    (CMD)    in
Fig.~\ref{fig:cmd}, constructed by assigning the parallaxes of primary
targets  to their  companion candidates.   The selection  criterion is
intentionally ``soft''  because some physical companions  are known to
deviate from the MS \citep[see discussion of Hipparcos dwarfs with
  deviant colors in][]{Koen10}.  The  nature of companions already listed in the
WDS  can be  established  with existing  data,  so most  of them  were
dropped  from  the  observing  program,  leaving 136  targets  in  the
list.  However,  known  companions  are included  in  the  statistical
analysis.

%-------------------------------------------------------------
\section{Observations and data reduction}
\label{sec:data}

\begin{figure}[ht]
\epsscale{1.0}
\plotone{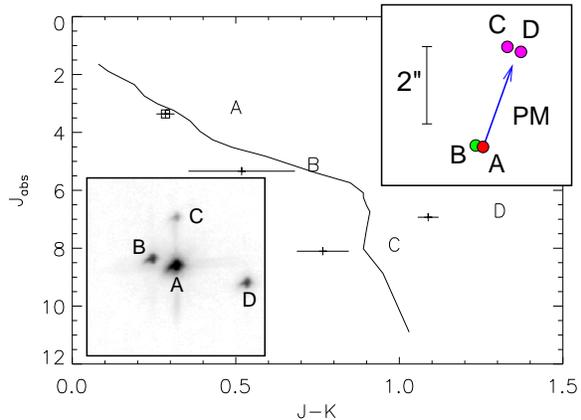}
\caption{Example of the data and their interpretation. There are 3 PSC
  companions  around  HIP~25662  with  separations  from  5\farcs5  to
  16\farcs9, labeled B,  C, and D respectively. The  location of these
  companions on the $(J_{\rm abs}, J-K)$ CMD is shown.  The lower-left
  insert is a fragment of  the combined $K_s$ image with 4-s exposure.
  The upper-right  insert shows the displacement of  the companions B,
  C, D relative  to their PSC positions and  the displacement produced
  by  the  proper motion  in  10 years  (arrow).  The  companion B  is
  physical, C and D are optical.
 \label{fig:example}}
\end{figure}

\subsection{Imaging data}

Simultaneous images in the visible and IR were taken at the CTIO 1.3-m
telescope (the  telescope used in  the 2MASS survey) with  the ANDICAM
instrument\footnote{http://www.astronomy.ohio-state.edu/ANDICAM/detectors.html}
\citep{ANDICAM}.  The  visual channel has  a CCD with  $1024^2$ binned
pixels of 0\farcs369 size, the IR channel has $512^2$ binned pixels of
0\farcs274.

Observations  were  carried out  in  February-March  2010 (2010.09  to
2010.19), in service  mode.   Five 4-s
images of each target were taken  in the $K_s$ band with dithers of 40
pixels  ($\sim 11\farcs6$),  followed  by five  30-s dithered  images.
Simultaneously, several  $V$-band images  with 2-s and  30-s exposures
were  recorded.  Most  results reported  here were  obtained  from the
short-exposure images.

There  are several problems  in the  data. HIP  10710 was  not pointed
correctly, HIP  59250, 60155, 60251,  60337, 61298, 61608,  73764 were
not observed  at all.  For 7  other targets (HIP  41620, 44579, 44777,
47312,  52676,  54285,  75839)  the short-exposure  sequence  was  not
executed,  leaving  only  30-s  exposures. The  median  image  quality
(full-width at  half maximum  from Gaussian fits)  is 0\farcs9  in the
$K$-band and 1\farcs1 in the $V$-band, 50\% of the data is within $\pm
10\%$  from  the  median  image  size.   Some  images  are  elongated,
especially in the $K$-band (25\% have ellipticity larger than 0.27).

Reduced images were retrieved from  the SMARTS data center at the Yale
University (the CCD images  are bias-subtracted and corrected for flat
field).  Dithered  IR images were  combined in the standard  way.  The
sky image was  calculated as a median over  dithered frames. Then each
sky-subtracted frame was  shifted to match the first  frame (the shift
was  determined by  cross-correlation).  The  re-centered  frames were
then median-combined.  Figure~\ref{fig:example} contains an example of
the combined  $K_s$ image with  3 companions around HIP~25662,  one of
which is  physical. This  star has a  known CPM companion  LDS~6186 at
99\farcs4   and   is  a   spectroscopic   binary   with  3.9y   period
\citep{Vogt02}. The  newly found companion makes this  system at least
quadruple.

\subsection{Relative astrometry and photometry}

Accurate relative  positions and  intensity ratios were  determined by
fitting the  secondary companion image with  the Point-Spread Function
(PSF)  of the primary.   Additional freedom  to treat  difficult cases
(blended or  slightly saturated images)  is provided by the  choice of
the inner and  outer radii for the fits.  Usually  the inner radius is
zero, the outer one is 6  pixels.  Even for saturated images, the fits
recover relative coordinates quite  well and give under-estimated, but
still useful  magnitude differences.   The results are  stable against
changes in  the fit  radius.  For close  companions, the  PSFs overlap,
leading to under-estimation of  the magnitude difference $\Delta m$ and
over-estimation of the separation $\rho$.

Astrometric calibration  of two CCD frames (crowded  fields around HIP
76572 and 36414) was done  by J.~Subasavage by referencing them to the
PSC. He  found the  pixel scale of  0\farcs3715 and angular  offset of
$1.5^\circ$ (to  be added to  the measured angles).  The  $K_s$ frames
were calibrated by comparing the relative companion positions with the
$V$-band  measurements.   The nominal  pixel  scale  of 0\farcs274  is
confirmed,  $-2.3^\circ$  has  to  be  added to  the  position  angles
measured in the IR images. Using these calibration parameters, I find
for  22 pairs  with  $\Delta  V <4$  the  average difference  $\langle
\theta_V  - \theta_K  \rangle  = 0.02^\circ  \pm  0.3^\circ$. The  rms
scatter between $V$ and $K_s$ positions is 0\farcs26 in separation and
$1.2^\circ$ in position angle.  These numbers estimate the measurement
errors,  dominated by  systematic effects  rather than  by  the random
noise.

\begin{figure}[ht]
\epsscale{1.0}
\plotone{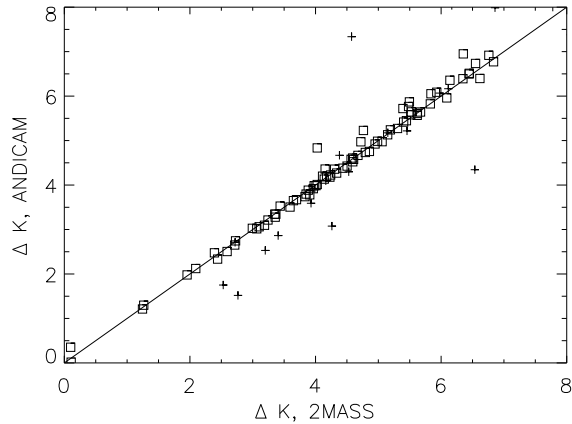}
\caption{Comparison of magnitude differences  in the $K_s$ band listed
  in the PSC with those measured  by ANDICAM for $\rho > 8''$.  The 80
  good-quality points (unsaturated, PSC quality flag A) are plotted as
  squares, the remaining 22 points as crosses.  \label{fig:dm}}
\end{figure}

Figure~\ref{fig:dm} compares our  differential photometry in the $K_s$
band with the PSC photometry  for separations $\rho > 8''$.  There are
only 12 cases  where the two $\Delta K$ measures  differ by more than
$0.5^m$; the robust estimate of the rms difference is $0.095^m$.

Relative astrometry and photometry  of candidate binaries is listed in
Table~1.    It  contains   the   {\it  Hipparcos}   number  and   pair
identification.   Then follow  the separation  $\rho$,  position angle
$\theta$ and  magnitude difference $\Delta K_s$ derived  from the PSC,
with   the  PSC   quality   flag  Q   for   the  $K$-band   photometry
\citep{2MASS}. The  remaining columns give $(\rho,  \theta, \Delta m)$
measured in  2010.14 with ANDICAM in  the $K_s$ and  $V$ bands.  Cases
where primary  components were saturated  are marked by the  flag {\rm
  S}.  For the  most part, I use measurements  from the short-exposure
images.

\subsection{New sub-systems?}

\begin{figure}[ht]
\epsscale{1.0}
\plotone{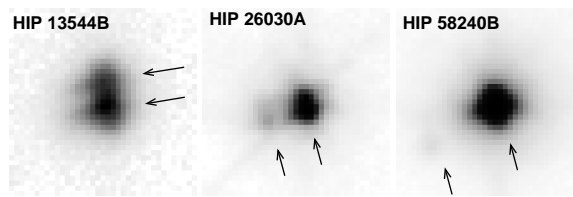}
\caption{Mosaic of three partially resolved pairs in $K_s$ band. Each
  square is $10''$ across, North is up and East left.
 \label{fig:subsyst}}
\end{figure}

Some objects  in the ANDICAM  images are resolved  close pairs. I  do not
discuss   such   pairs   for    optical   companions   and   show   in
Fig.~\ref{fig:subsyst}  only  three  most  obvious cases  of  resolved
physical  companions.   Measurements  of  the relative  positions  and
magnitude  differences  in  these  close pairs  are  only  approximate
because  the PSFs overlap.

{\it  HIP 13544BC}  is a  known pair  A~2341, making  a  triple system
together with the  main target. ANDICAM resolves BC  into nearly equal
stars at $1\farcs3, \; 4^\circ, \; \Delta K \sim 0.6$.

{\it HIP 26030A}  has a faint satellite at  $1\farcs6, \; 91^\circ, \;
\Delta K = 2.7$. The new companion could be optical because this field
is  crowded, $N^*  = 57$;  there  are 4  other stars  in the  $K$-band
ANDICAM images and many more in the CCD frames.

{\it HIP 58240B}  is in fact a pair  Ba,Bb with parameters $3\farcs6,
\; 116^\circ, \; \Delta K = 5.8$. Further observations will help to
determine the status of this pair in a moderately crowded field with
$N^* = 34$. 

%-------------------------------------------------------------
\section{Results}
\label{sec:res}

%\subsection{Assembling the data}

\begin{figure}[ht]
\epsscale{1.0}
\plotone{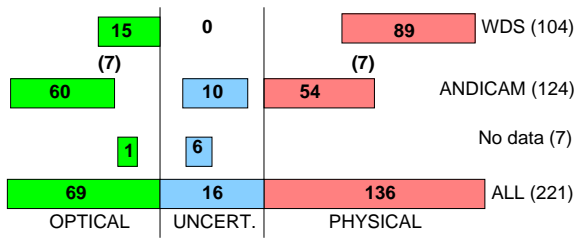}
\caption{Distribution of companions over various categories. 
 \label{fig:count}}
\end{figure}

The sample of binary stars is assembled by merging companions found in
the PSC with  existing data from the  WDS in the same part  of the sky
(RA from  $2^h$ to $18^h$,  $\delta < +20^\circ$).  The  separation of
WDS  companions considered  here  is from  $5''$  to $20''$,  although
closer  companions found  in the  PSC are  included in  the  sample as
well. The WDS binaries in  crowded fields ($N^* > 100$) are excluded
to avoid statistical bias.  These selection criteria are satisfied for
1913 primary stars  in the {\it Nsample}.  I  ignore 18 WDS companions
not detected by  the PSC for various reasons  (e.g.  primary component
too bright or partially resolved)  and exclude 15 PSC companions which
are  likely artifacts  (not  detected in  all  bands by  2MASS or  not
confirmed  here).   There  are  221 binary  companions  or  candidates
remaining.

All companions are classified into three groups: true (physical),
background stars (optical), and uncertain. Three criteria are used.
\begin{enumerate}
\item
Astrometry,  i.e. constancy  of the  relative companion  position over
time. For  the known pairs,  the first-epoch positions are  taken from
the WDS, the  last-epoch from the PSC. The  new candidate binaries are
checked by comparing the PSC positions with the second-epoch 2010 data. The
change of the relative position is compared to the reflected PM
of the primary component (Fig.~\ref{fig:example}). 

\item
Companion location in the $(J_{\rm abs}, J-K)$ CMD with respect to the
MS line \citep{Lang92}, using  the distance modulus based on $\pi_{\rm
  HIP}$. % (Fig.~\ref{fig:jk}).  
The photometry comes from the PSC.

\item
Companion    location    in    the    $(K_{\rm   abs},    V-K)$    CMD.
%(Fig.~\ref{fig:vk}). 
The $V$-magnitudes are taken from the WDS for known
pairs or  determined from the  $\Delta V$ measured here  for candidate
companions detected in the $V$-band.
\end{enumerate}

If random  and systematic  errors of the  data used in  these criteria
were  known, the  formal probability  of  passing the  tests could  be
computed.  However, such  a  formal approach  makes  little sense  for
various  reasons.  The astrometric  criterion (1)  is affected  by the
orbital  motion in  the wide  binaries, by  motions of  the components
caused by  un-detected inner sub-systems, and, mostly,  by the unknown
PMs  of  the  background  stars.   In  many  instances  the  available
astrometry  allows a  clear distinction  between optical  and physical
companions (Fig.~\ref{fig:example}), but  there remain marginal cases,
especially  for  targets with  small  PM  and  10-year time  coverage.
Similarly, the application of the  criteria (2) and (3) is affected by
errors in the photometry and by the possibility of physical companions
being located off the MS.

I  evaluated each  of the  3  criteria subjectively  on a  continuous
scale, assigning  negative values for optical  companions and positive
values  for  physical ones.   Larger  absolute  numbers mean  stronger
evidence,  zero   stands  for  a complete  lack   of  information.   The
combination of all 3 criteria  is resumed in the opticity flag $O$,
with $-2$ and  $-1$ for certain or almost  certain optical companions,
$+1$ and $+2$ for very likely and certain physical companions.

Figure~\ref{fig:count}  presents the  distribution of  companions over
the three categories (optical,  uncertain, physical) and over the data
sources  (WDS,  ANDICAM), indicating  partial  overlaps between  these
groups. The sample  contains 136 physical companions, 47  of which are
new (discovered  in the  PSC and confirmed  with ANDICAM).   This work
thus increases the  number of known physical companions  in the $(5'',
20'')$ separation range around nearby G-dwarfs by 55\%.

The results are summarized in  Table~2, one line per pair. The systems
are identified  by the  HIP numbers of  their primary  components (the
cases where secondaries have distinct HIP numbers are indicated in the
notes,  Table~\ref{tab:notes}).   The  $V$-magnitudes of  the  primary
components are taken  from the Tycho catalog \citep{HIP},  the $J$ and
$K_s$  magnitudes  from the  PSC.   For  each  primary target  I  list
$\pi_{\rm HIP}$ and $N^*$.  For the secondary and tertiary components,
the separation $\rho$ and position angle $\theta$ based on the PSC are
given.  The  $V$-band photometry of  the known secondaries  comes from
the WDS, for  some new candidates -- from  our measurements of $\Delta
V$.  The  flag W=1 is  set if the companion  is known (e.g. found  in the
WDS), flag  A=1 means  that the companion  was observed  with ANDICAM.
The last  column contains the  opticity flag $O$.  Table~2  also lists
masses of  the stars estimated  from their absolute $K$  magnitudes by
relations  of  \citet{HM93}; obviously,  for  optical companions  such
estimates are meaningless.

More information  (nature of  the sub-systems etc.)   is given  in the
notes  (Table~\ref{tab:notes}).   In   the  following,  I  ignore  the
sub-systems, considering each  component of a wide binary  as a single
star even when it  is known to be a pair. This  survey has produced 15
new multiple  systems (HIP 11417,  11537, 12764, 25148,  25662, 34212,
36832,  39999,  41871,  47312,  52145, 54366,  56282,  68507,  84866).
Spectroscopic companions to  primary targets discovered by \citet{N04}
are  marked as  N04 in  the notes  together with  the range  of radial
velocity  variation   $\Delta  RV$  in   km/s.   Astrometric  binaries
discovered by  \citet{Makarov05} are reported.   About a third  of the
new    wide   binaries    are    actually   higher-order    multiples;
\citet{Makarov08}  encounter a similarly  high fraction  of multiples,
25\%, in their sample  of CPM pairs.  I give in the  notes the PMs (in
mas/yr) and  $V$-band photometry from the  NOMAD catalog \citep{NOMAD}
for companions which are found there.  However, the sources and errors
of  the  PMs in  NOMAD  are  not known,  and  in  many instances  they
contradict  my findings  (discrepant  PMs for  physical companions  or
matching PMs for optical companions).  Therefore, the information from
NOMAD is ignored in deciding the companion status. The UCAC2 catalog
\citep{UCAC2} is not deep enough ($R < 16$) to reach the faintest
secondaries. 

\begin{figure}[ht]
\epsscale{1.0}
\plotone{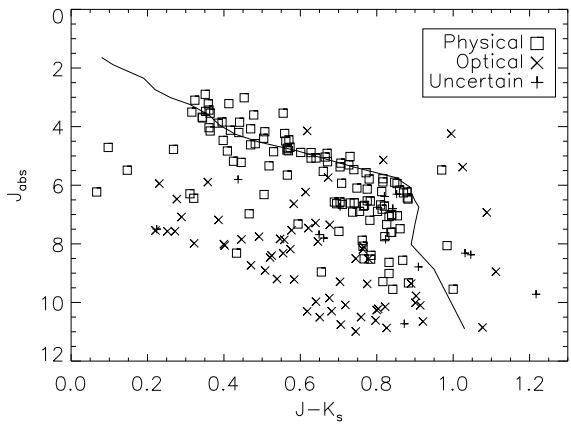}
\caption{$(J_{\rm abs}, J-K)$ CMD of  the secondary companions with $\rho  > 5''$. Standard
  MS according to \citet{Lang92} is shown.
 \label{fig:jk}}
\end{figure}

\begin{figure}[ht]
\epsscale{1.0}
\plotone{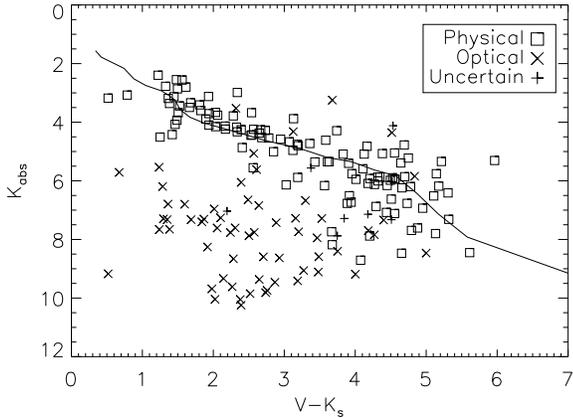}
\caption{$(K_{\rm  abs}, V-K)$  CMD of  the secondary  companions with  $\rho > 5''$. The line shows the MS.
 \label{fig:vk}}
\end{figure}

Figure~\ref{fig:jk} is the  $(J_{\rm abs}, J-K)$ CMD based  on the PSC
photometry.  Most  physical companions concentrate around  the MS. The
spread of  the points is large  and there are  several true companions
below the MS, some of  those quite bright.  These outliers mostly have
large photometric  errors in the PSC.  However, there is  a real $\sim
0.2^m$ spread in  the $J-K$ colors of physical  companions, as follows
e.g.   from the  work  of \citet{HM93}.   Most  optical and  uncertain
companions  are  located  below   the  MS.   The  companion  selection
criterion (\ref{eq:crit})  leaves a convenient  margin, admitting many
optical candidates but not  rejecting physical companions with deviant
colors.

The  $(K_{\rm abs}, V-K)$  CMD is  plotted in  Fig.~\ref{fig:vk}.  The
$V-K$  color  has  a  larger  sensitivity  to  effective  temperature,
compared to $J-K$, and therefore discriminates better between physical
and  optical   companions.   The   number  of  physical   outliers  is
correspondingly  less.   However,  faint  and red  low-mass  companion
candidates were not detected in the $V$ band with ANDICAM, leaving the
lower right  corner of the  CMD empty.  Few physical  companions below
the MS are explained by  largely under-estimated $\Delta V$ in ANDICAM
images  with  saturated  primary  stars (the  companions  then  appear
bluer).

%-------------------------------------------------------------
\section{Statistical analysis}
\label{sec:stat}

\subsection{Detection limit in the PSC}

\begin{figure}[ht]
\epsscale{1.0}
\plotone{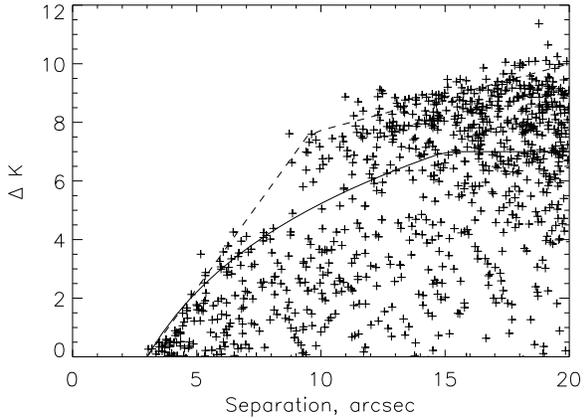}
\caption{Magnitude difference  $\Delta K$  vs. separation for 982 PSC
  companions with valid photometry. The dashed and full lines indicate
  the realistic and pessimistic detection thresholds, respectively.
 \label{fig:det}}
\end{figure}

Detection of faint  sources in the PSC is  complicated by the presence
of nearby bright stars (primary  targets). It is important to estimate
this bias.   Some sources are detected  only in the $K$  band, most of
them are artifacts.  They are, however, not removed from the candidate
list,  and some  turn  out  to be  genuine  physical companions.   For
evaluating the PSC detection bias,  I select companions with valid PSC
photometry in all  three bands $JHK_s$ within $20''$  from the targets
and plot  the magnitude difference with primary  components $\Delta K$
vs.   separation  $\rho$   in  Fig.~\ref{fig:det}.   The  dashed  line
approximates the upper envelope by two segments.

A primary target of 1\,$M_\odot$ has $K_{\rm abs} \approx 3.1$, so the
restriction on  the absolute magnitude  $J_{\rm abs} < 11$  imposed on
the  candidates translates  to $\Delta  K  < 7$.  The actual  physical
companions  obey this  restriction  and delineate  a more  pessimistic
threshold
\begin{equation}
\Delta K_{\rm lim} = A \log_{10} (\rho/3'') 
\label{eq:det2}
\end{equation}
with $A = 10$ (solid  line in Fig.~\ref{fig:det}). 
Statistical interpretation of the data depends on the adopted
detection limit.

%-------------------------------------------------------------
\subsection{Distribution of the mass ratio}
\label{sec:q}

\begin{figure}[ht]
\epsscale{1.0}
\plotone{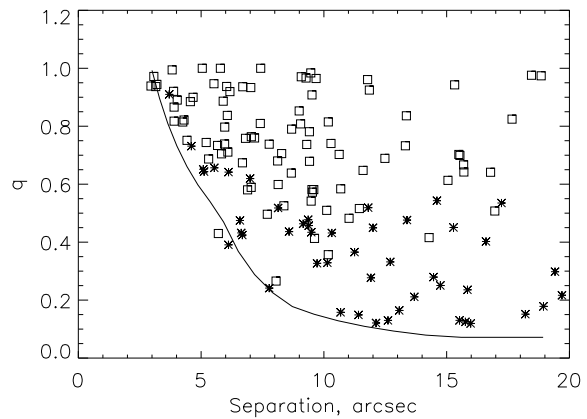}
\caption{Mass ratios of binaries vs. separation. Squares -- previously
  known  systems, asterisks  -- new  systems from  the PSC.   The line
  shows the detection limit (\ref{eq:det2}) with $A=10$.
 \label{fig:q-sep}}
\end{figure}

The statistical  analysis is performed  on 120 physical  binaries with
$\rho \ge 5''$ selected by  the criterion $O \ge 1$ (Table~2).  Masses
of  the  primary  and  secondary  components,  $M_1$  and  $M_2$,  are
estimated  from  their absolute  magnitudes  $K_{\rm  abs}$ using  the
relation from  \citep{HM93}. For 3893 targets of  {\em Nsample} common
with the  survey of  \citet{N04} I find  a good agreement  with their
mass estimates, to  better than $\pm 10$\%.  For  the binaries studied
here, $M_1$  ranges from 0.68 to  1.52 $M_\odot$, the  median $M_1$ is
1.03\,$M_\odot$.

Figure~\ref{fig:q-sep}   plots  the   mass  ratio   $q  =   M_2/  M_1$
vs. separation  $\rho$.  The estimated PSC detection  limit in $\Delta
K$ is translated into $q$ assuming $M_1 = 1\,M_\odot$. At first sight,
the points  are distributed  uniformly in $q$  in the space  above the
detection limit. There appears to be a slight preference for small $q$
at $\rho  >12''$ and a tendency  to large $q$  at smaller separations.
The  actual detection  limit of  the PSC  remains  somewhat uncertain,
affecting the statistical analysis.   I tried several assumptions. The
results reported below correspond  to the hard detection limit $A=10$,
supposing  that all  fainter companions  are  missed in  the PSC,  all
brighter ones are detected.

Considering  that  the  companion  detection  ``depth''  is  a  strong
function of  $\rho$, I study  the mass ratio distribution  $f(q)$ for
pairs with $\rho \ge \rho_{\rm min}$, for different cutoffs $\rho_{\rm
  min}$.   The assumption  is  made that  separations are  distributed
according  to  the  \"Opik's  law  (constant  in  $\log  \rho$),  well
established for  separations around $10^3$\,AU  \citep{Poveda04}.  The
companion frequency $\epsilon$ is normalized per decade of separation,
i.e.   divided by  $\log (20''/\rho_{\rm  min}$), and  referred  to the
total number of  {\it Nsample} targets in the  surveyed portion of the
sky, $N=1913$. 

A power-law model of mass-ratio distribution
\begin{equation}
f(q) = \epsilon (\beta +1) q^\beta 
\label{eq:fq}
\end{equation}
is  frequently  used  in  the literature  \citep[e.g.][]{MH09}.   Here
$\epsilon$ is the total companion fraction, $\beta$ is the power index
(slope). These two parameters  are related because correction for the
missed low-mass companions depends on $\beta$. 

\begin{figure}[ht]
\epsscale{1.0}
\plotone{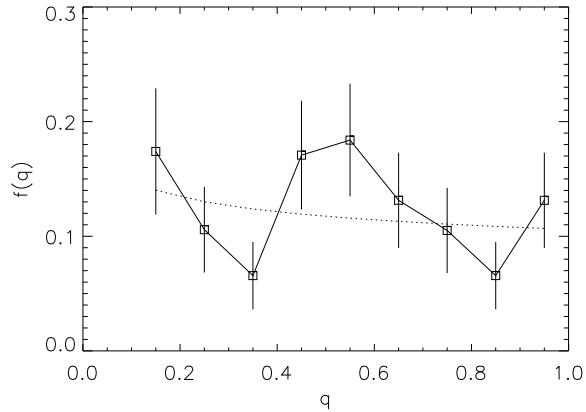} %qhist8a.ps
\caption{Mass-ratio histogram  of 83 physical companions  with $\rho >
  8''$  in 0.1-wide  bins,  corrected for  incomplete detection.   The
  dashed line shows  a power-law fit. The integral  of $f(q) {\rm d}q$
  gives the companion frequency per decade of separation.
 \label{fig:qhist8}}
\end{figure}

\begin{figure}[ht]
\epsscale{1.0}
\plotone{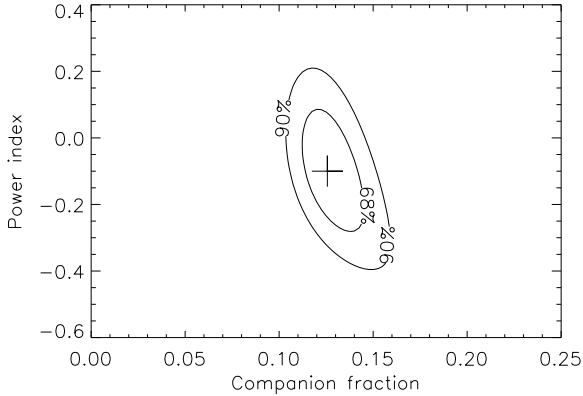} %ml8a.ps
\caption{Confidence  area in the  $(\epsilon, \beta)$  parameter space
  for the ML solution with $\rho_{\rm min} = 8''$ and  $A=10$.
 \label{fig:ml8}}
\end{figure}

A histogram of the mass  ratios is shown in Fig.~\ref{fig:qhist8}. The
number  of  companions  in  each  bin  is  increased  to  account  for
incomplete detections (divided by the average detection probability in
each $q$-bin  in the separation  interval $\rho_{\rm min},  20''$) and
normalized  per  decade  of  separation.   The first  bin  with  large
incompleteness is avoided, the remaining  data are fitted by a line in
the  log-log  coordinates (a  power  law)  with  $\beta =  -0.15$  and
$\epsilon =  0.13$.  The power-law distribution  describes the observed
histogram adequately.

Alternatively,  the model  (\ref{eq:fq})  can be  fitted  to the  data
directly  by the  maximum likelihood  (ML)  method, as  done e.g.   in
\citep{NICI}.  The  detection limit is  included in the  analysis. The
likelihood   function  is   usually  interpreted   as   a  probability
distribution in  the parameter space,  enabling the definition  of the
confidence area  and a better  visualization of the  mutual dependence
between parameters  \citep{NumRec}.  On the other hand,  the ML method
requires a parametrization, in this case (\ref{eq:fq}).

The  results  of the  ML  analysis  depend  on the  assumed  companion
detection  limit and on  the separation  cutoff $\rho_{\rm  min}$.  By
adopting  a  more  strict  (pessimistic) detection  limit,  I  obtain
smaller values of $\beta$ and  less dependence of $f(q)$ on $\rho_{\rm
  min}$.  The  derived companion  fraction  $\epsilon$ remains  nearly
constant.  It is  expected that $f(q)$ should not  depend on $\rho$ in
the      relatively      small      interval      considered      here
\citep[e.g.][]{Raghavan10}.
After trying  several detection limit  models, I finally use  a sharp
threshold with $A=10$.  Please, keep  in mind that the ML results will
be somewhat different for other detection models.

\begin{figure}[ht]
\epsscale{1.0}
\plotone{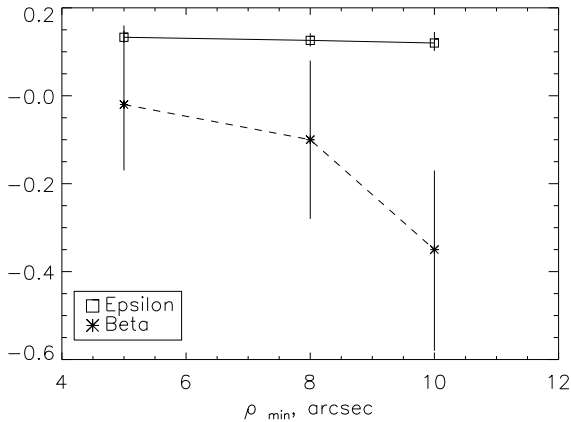} %mlplot.ps
\caption{Dependence of the best-fit parameters  $(\epsilon, \beta)$
  on the  $\rho_{\rm min}$ for $A=10$.
 \label{fig:ml}}
\end{figure}

Figure~\ref{fig:ml8} shows a representative result of the ML procedure
for $\rho_{\rm  min} = 8''$  (83 companions). The  best-fit parameters
are  $\epsilon  =  0.13$ and  $\beta  =  -0.10$,  the lines  show  the
confidence areas  in the parameter space corresponding  to 68\% (``one
sigma'')  and  90\%.   The  confidence  area is  skewed,  therefore  I
obtained the standard errors of parameter estimates by integrating the
probability  distribution over other  coordinate.  Figure~\ref{fig:ml}
shows  the dependence  of  the best-fit  parameters  and their  formal
$\pm \sigma$  errors  on   $\rho_{\rm  min}$.   The  companion  fraction
$\epsilon$ is rather stable against $\rho_{\rm min}$, ranging between
0.15 and 0.13,  but the power-law index $\beta$  becomes more negative
at  larger separations,  in  agreement with  the  trend noticeable  in
Fig.~\ref{fig:q-sep}.

\subsection{Alternative analysis}

Considering the uncertainty  of the PSC detection limit  for faint and
close  companions, I  carried out  an alternative  analysis  on wider
companions between $10''$ and $30''$. This time, the PSC data over the
whole sky are  used.  Only candidates with $J_{\rm  abs} < 11$ (bright
enough to be stellar  companions) and reliable PSC photometry (quality
flags  A--D or E  in $JHK_s$  bands) are  retained. Scaling  the total
number of  such candidates within $2.5'$  radius to the  $10'' - 30''$
annulus,  I expect to  find 2502  and actually  find 2115.   There is
still  a small  deficiency  of candidates  indicative  of a  potential
detection bias. If all companions without constraints on $J_{\rm abs}$
are selected,  the expected number  is twice the actual  number (10177
and  5299, respectively). Therefore,  the PSC  magnitude limit  in the
vicinity of bright stars is indeed not as deep as in the field, even
at $10''$ separation.

\begin{figure}[ht]
\epsscale{1.0}
\plotone{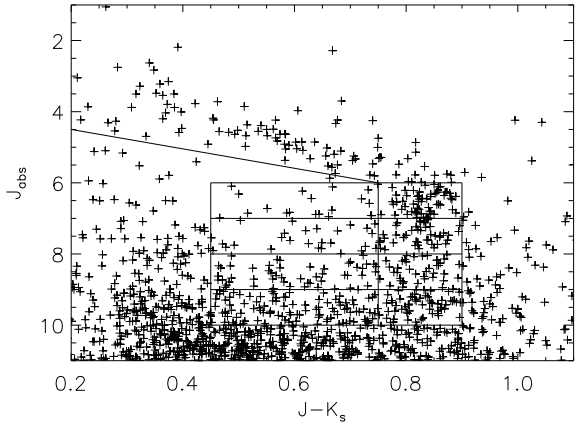}  %cmdbox.ps
\caption{The $(J_{\rm abs}, J-K)$ CMD of  PSC companions between
  $10''$ and $30''$ with reliable photometry. The lines indicate box
  areas used for statistical analysis. 
 \label{fig:box}}
\end{figure}

Physical  companions  can  be  distinguished  from  the  optical  ones
statistically, if not individually. To ease this task, I restrict the
study to 4473 primaries where the  number of stars with $J_{\rm abs} <
11$  within $2.5'$  radius  is less  than  100. The  CMD  of the  1470
companions  in  the  $(10'',  30'')$  separation  range  is  shown  in
Fig.~\ref{fig:box}.  Bright companions clearly concentrate towards the
MS, merging progressively with background stars at fainter magnitudes.
The  separation  distributions  of  bright and  faint  companions  are
different,  the former  following  the \"Opik's  law $f(\rho)  \propto
1/\rho$ and the latter being proportional to $\rho^2$.

\begin{figure}[ht]
\epsscale{1.0}
\plotone{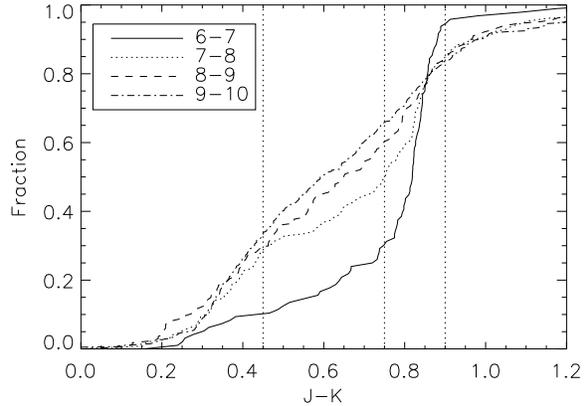}  %jkhist.ps
\caption{Cumulative  distributions  of the  $J-K$  color in  different
  intervals of $J_{\rm abs}$,  shown in the legend.
 \label{fig:jkhist}}
\end{figure}

\begin{figure}[ht]
\epsscale{1.0}
\plotone{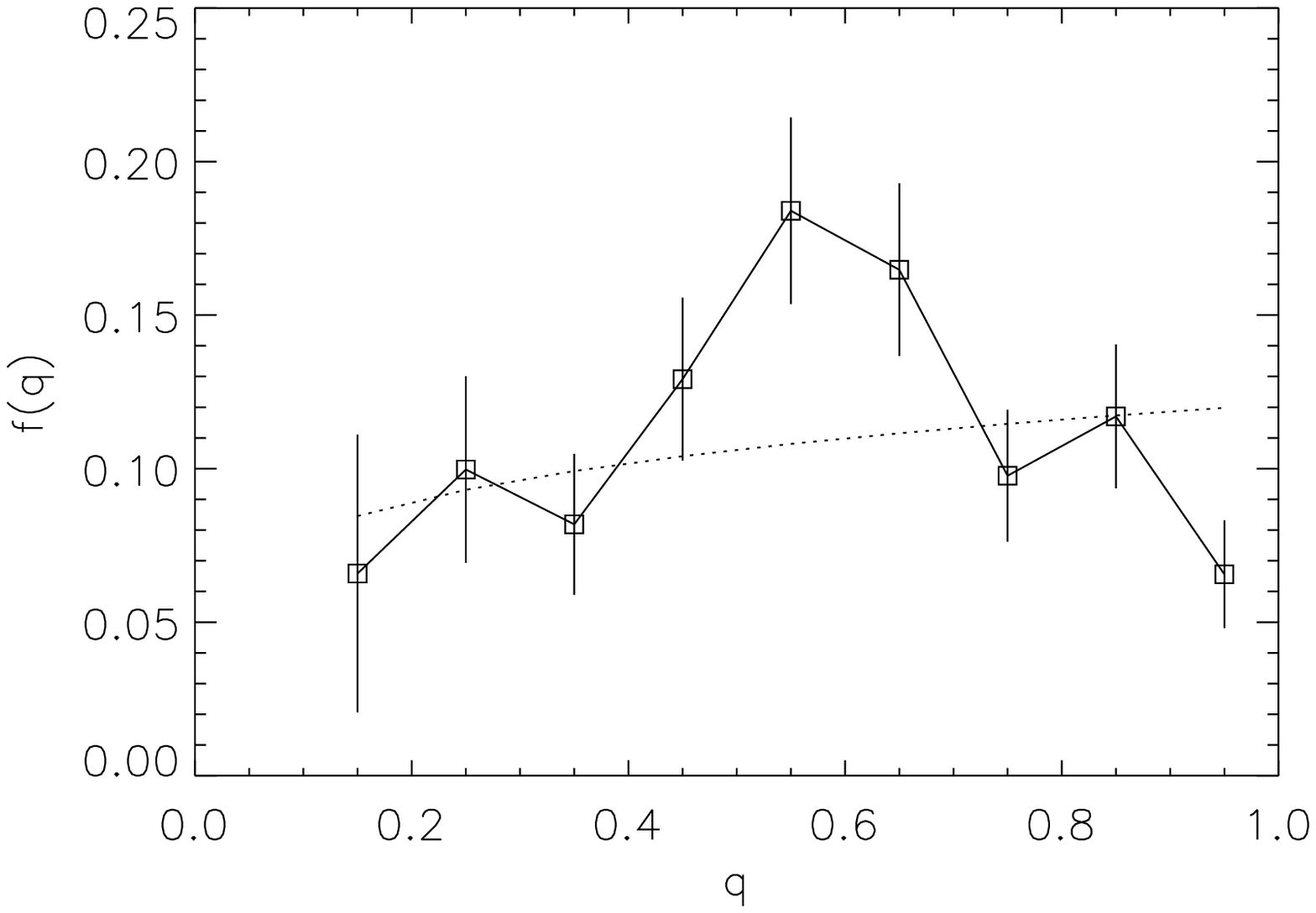}  %f-qa.ps
\caption{Histogram of mass ratio  for PSC companions between $10''$
  and  $30''$  corrected  for  contamination.  The dotted  line  is  a
  power-law fit with $\beta = 0.19$.
 \label{fig:f-q}}
\end{figure}

Cumulative  distributions of  the  $J-K$ color  in  four intervals  of
$J_{\rm abs}$ are plotted in Fig.~\ref{fig:jkhist}. The vertical lines
show the  color range for the  physical companions $0.75 <  J-K < 0.9$
(P)  and the  comparison range  $0.45  < J-K  < 0.75$  (O) where  only
optical companions  are expected.   These intervals correspond  to the
boxes  in Fig.~\ref{fig:box}.   Obviously, most  companions with  $6 <
J_{\rm abs} <7$ are physical,  but the companion density in the ranges
P  and O  becomes more  similar and  eventually equal  with increasing
$J_{\rm abs}$.   This gives a way  to estimate the  number of physical
companions $n_c$ in each interval of $J_{\rm abs}$ as
\begin{equation}
n_c = n_p - 0.5 n_o = \alpha n_p, \;\;\; \alpha = 1 - n_o/(2 n_p).
\label{eq:nc}
\end{equation}
This formula assumes that optical companions are uniformly distributed
in $J-K$, that physical companions are concentrated in the P-interval,
and  that the O-interval is  two times  larger.  The  contamination factor
$\alpha$  decreases from one  to nearly  zero with  increasing $J_{\rm
  abs}$.   The rms  error of  $n_c$  estimate is,  obviously, $(n_p  +
n_c/4)^{1/2}$. Table~\ref{tab:count} lists the  companion
counts.

The fraction  $\alpha$ of companions  in the P-interval and  above the
tilted line in Fig.~\ref{fig:box} are physical.  Masses of the primary
stars and companion candidates  are estimated from their $K_{\rm abs}$
magnitudes using  the relation from \citep{HM93}, the  mass ratios $q$
are  calculated  and the  distribution  of  $q$  in 0.1-wide  bins  is
constructed,   applying   the   factors   $\alpha$  to   correct   for
contamination.    These   factors    are   modeled   after   data   in
Table~\ref{tab:count} as $\alpha(q) \approx  1 - 1.5 (1- q)^{4}$; they
become significant  for $q<0.5$, otherwise  are close to 1.   The last
column  of Table~\ref{tab:count} gives  approximate mass  ratios $q_0$
corresponding to the centers of  the $J_{\rm abs}$ intervals for $M_1 =
1\,M_\odot$.

Figure~\ref{fig:f-q} shows  the resulting histogram. The  first bin is
avoided because it  has a large error and  a large $\alpha$-correction
The values of $f(q)$ are normalized per decade of separation, assuming
the \"Opik's law.   The dotted line shows a linear  fit in the log-log
coordinates, i.e.  a power law.  The exponent is $\beta = 0.19$.  The
sum of $f(q) {\rm d}q$ over all  bins except the first one leads to
$\epsilon = 0.10$. 

In   comparison   with   the   histogram   in   Fig.~\ref{fig:qhist8},
Fig.~\ref{fig:f-q}  shows a smaller  fraction of  low-mass companions,
presumably  because I  do not  correct here  for  incomplete detection.
This also explains why  the estimated companion fraction $\epsilon$ is
slightly  less  than  in  the  sub-section~\ref{sec:q}.  However,  the
companions  considered here  are  wider, hence  less  affected by  the
bright glow of the primary targets.  The $f(q)$ distributions obtained
by two different approaches on different (although overlapping) samples
of companions are  close to each other, except for  the first bins at
$q <0.3$.  Therefore, the previously adopted detection threshold seems
to be a reasonable choice.  Despite the remaining uncertainties, it is
clear that $f(q)$ is nearly  uniform and does not rise dramatically at
$q < 0.3$.

An interesting conclusion  from this study is that  the PSC companions
within $30''$  from the  targets with $J_{\rm  abs} < 7$  and suitable
$J-K$ colors are  physical with a probability of  $\ge 84$\%. Many of
those  companions not  listed in  the  WDS are  strong candidates  for
checking their status with second-epoch imagery. This will lead to the
census  of visual  companions around  {\it Nsample}  G-dwarfs complete
down to $q \sim 0.2$.

%-------------------------------------------------------------
\section{Discussion}
\label{sec:disc}

The    sample   of   low-mass    visual   companions    assembled   in
Section~\ref{sec:res}  can still be  biased. On  one hand,  some known
pairs with  large $q$ are  excluded because they  do not appear  in the
PSC. On  the other hand,  some low-mass companions are  missed because
their physical status could not yet be confirmed.  The total number of
missed companions  in each of these  two groups is  $\sim 10\%$ of
the total number of binaries.  Their inclusion would slightly increase
the estimated  companion fraction $\epsilon$ and  would affect $f(q)$,
although  not  dramatically. The  largest  uncertainty  of the  derived
$f(q)$  is related  to the  adopted model  of the  companion detection
limit in the PSC.

I intentionally  ignored inner  sub-systems. The masses  are estimated
from  the total  light in  the $K$  band. For  unresolved sub-systems,
these estimates  are closer  to the mass  of their  primary components
than to the total mass.   As the knowledge of sub-systems is currently
very incomplete (especially  for the secondary companions), correction
for them  could be  only partial.  The  $f(q)$ derived here  remains a
preliminary estimate  until the multiplicity  survey of the  sample is
done  to  enable  a  reasonably  comprehensive account  of  the  inner
sub-systems.

The  distribution  of  the  mass  ratio  $q$  of  wide  companions  to
solar-type dwarfs is found here  to be nearly uniform.  These binaries
have typical  projected separations of $10^3$\,AU  and orbital periods
on the order  of $10^7$ days.  \citet{MH09} find  a similar mass-ratio
distribution with $\beta  = -0.39 \pm 0.36$ in a  sample of 30 closer
companions  to  solar-type stars  surveyed  with  adaptive optics.   A
mass-ratio  distribution  with  $\beta   \sim  -0.5$  was  derived  by
\citet{ST02}  for visual  companions to  B-type stars  in the  Sco OB2
association,  while \citet{Kraus08}  found  a nearly  flat $f(q)$  for
low-mass visual  binaries in this association.  Anyway, the conclusion
of  \citet{DM91} that  masses  of wide  companions  to G-dwarfs  match
random selection from the initial mass function is now firmly refuted.

Possible  dependence  of  $f(q)$  on  the  orbital  period  remains  a
debatable subject.  \citet{DM91} found  no convincing evidences of any
such  dependence in  their  22-pc sample  of  164 G-dwarfs.   However,
modern study by \citet{Raghavan10}  of the 25-pc sample containing 454
dwarf stars does  show a clear trend to smaller  $q$ in wide binaries,
while  short-period  pairs prefer  equal-mass  components (cf.   their
Fig.~17).  They find that  the mass-ratio distribution integrated over
all periods  is remarkably flat and  declines at $q <  0.1$.  It seems
well established  that for spectroscopic binaries  with orbital periods
below   $10^3$\,d   the   $f(q)$   is  nearly   flat   \citep{Mazeh03,
  Halbwachs03}. The  mass-ratio distribution in  the inner sub-systems
of multiple stars  with solar-type primaries also appears  to be flat,
to the best of our knowledge \citep{NICI}.

The frequency  of companions to solar-type stars  with orbital periods
on the order of $10^7$\,d is around 0.08 per decade of period, or 0.12
per  decade   of  separation   according  to  both   \citet{DM91}  and
\citet{Raghavan10}.  The  companion frequency estimate  obtained here,
$0.13  \pm  0.015$,  is   essentially  the  same,  but  more  accurate
statistically, owing to the larger number of binaries.

This work shows  the great potential of the  2MASS PSC for discovering
new low-mass  companions.  Further observations are  needed to confirm
candidate  companions   at  larger  separations  and   in  other,  yet
un-surveyed, parts  of the  sky in order  to reach complete  census of
wide binaries in {\it  Nsample}.  Combination of this information with
spectroscopic and adaptive-optics surveys of the same sample will open
a unique possibility to  obtain comprehensive statistics of binary and
multiple  systems with  unprecedented accuracy  and detail.   This, in
turn,  will  advance  our  understanding  of star  formation  and  our
origins.

%-------------------------------------------------------------
%-------------------------------------------------------------
%-------------------------------------------------------------
%-------------------------------------------------------------

\acknowledgments I thank  SMARTS observers J.~Vasquez, A.~Miranda, and
J.~Espinosa for  making observations and  SMARTS data archive  at Yale
managed by  S.~W. Tourtellotte  for pre-processing and  delivering the
images. The help of J.~Subasavage with astrometric calibration is much
appreciated. This work used the  2MASS data products, WDS catalog, ADS
services, and SIMBAD.

{\it Facilities:} \facility{SMARTS}

%\input{table1.tex}

%\input{table2.tex}
%\input{table3.tex}
%\clearpage

\begin{deluxetable}{c l | cccc | cccc | cccc}                                                                                                  
\tabletypesize{\scriptsize}                                                                                                               
%\rotate                                                                                                                                  
\tablecaption{Companion measurements  (2010.14)   
\label{tab:res} }                                                                                                                         
\tablewidth{0pt}                                                                                                                          
\tablehead{
& & \multicolumn{4}{|c}{ 2MASS} & \multicolumn{4}{|c}{ANDICAM (K)} &  \multicolumn{4}{|c}{ANDICAM (V)} \\
HIP & Sys & $\rho$ &   $\theta$ & $\Delta K$  & Q 
& $\rho$ &   $\theta$ & $\Delta K$  & S 
& $\rho$ &   $\theta$ & $\Delta V$  & S \\ 
& &  $''$ & $^\circ$ &   mag & & $''$ &  $^\circ$ &   mag & & $''$ & $^\circ$  &   mag &  }
\startdata   
  9499& AB &  12.20 &  225.2 &   6.83 & A &  12.59 &  224.9 &   6.77 &   &  12.63 &  225.7 &   7.48 &   \\
 10621& AB &  18.84 &  202.6 &   5.93 & A &  18.39 &  202.2 &   6.09 &   &  18.43 &  203.0 &   7.64 &   \\
 11024& AB &  10.29 &   81.1 &   1.24 & A &  10.34 &   82.2 &   1.21 &   &  10.05 &   79.6 &   2.52 &   \\
 11417& AB &  19.41 &  297.7 &   3.85 & A &  19.82 &  297.6 &   3.80 &   &  19.48 &  297.2 &   6.81 &   \\
 11537& AB &   3.70 &  346.5 &   0.39 & D &   4.14 &  343.3 &   3.15 &   &   4.15 &  345.2 &   5.36 &   \\
 11909& AB &  15.76 &  116.9 &   6.11 & A &  15.83 &  117.1 &   6.17 & S &        &        &        &   \\
 12764& AB &   5.12 &  271.6 &   1.83 & D &   5.77 &  273.4 &   3.85 &   &        &        &        &   \\
 13544& AB &  18.45 &   23.8 &   0.10 & A &  18.05 &   22.3 &   0.35 &   &  18.29 &   23.6 &   0.83 &   \\
\enddata                                                                                                                                  
\end{deluxetable}                                                                                                                         

\begin{deluxetable}{ccc   rrrr | l cc rrrr cr} 
\tabletypesize{\scriptsize}                                                                                                               
%\rotate                                                                                                                                  
\tablecaption{Photometry and companion status       
\label{tab:ptm} }                                            
\tablewidth{0pt} 
 %HIP plx N* Vmag Jmag Kmag Mass | Comp rho theta  Vmag Jmag Kmag Mass WA O   ==> 16 items
\tablehead{
\multicolumn{7}{c}{Primary target} & \multicolumn{9}{|c}{Companion} \\
\colhead{HIP} & \colhead{$\pi_{\rm HIP}$} & \colhead{$N^*$} 
& $V$ & $J$ & $K_S$  & $M_1$ 
&   & \colhead{$\rho$} &   \colhead{$\theta$} 
&  \colhead{$V$} &   \colhead{$J$} & \colhead{$K_S$}  & \colhead{$M_2$}
& \colhead{WA} & \colhead{$O$} \\
 & mas & &    & & & $M_\odot$ & & $''$ & $^\circ$ & & & &  $M_\odot$ & & }
\startdata   
  9499&   15.3&   11&   8.07&   6.94&   6.58&   1.15& B&  12.20&  225&  15.55&  14.05&  13.41&   0.10& 01&   -2.0 \\
 10175&   24.4&   13&   8.66&   6.91&   6.58&   0.90& B&   6.04&  243&   9.13&   7.21&   6.85&   0.84& 10&    2.0 \\
 10305&   25.2&   14&   5.65&   5.05&   4.33&   1.52& B&  16.79&  233&   6.69&   6.54&   6.17&   0.97& 10&    2.0 \\
 10579&   17.2&    6&   9.24&   8.31&   7.97&   0.77& B&   6.70&  288&   9.66&   8.65&   8.24&   0.72& 10&    2.0 \\
 10621&   16.9&   13&   7.78&   6.85&   6.56&   1.09& B&  18.84&  202&  15.42&  13.08&  12.49&   0.13& 01&   -1.0 \\
 10710&   15.9&   13&   8.93&   7.72&   7.33&   0.94& B&  14.39&  312&    -  &  10.79&   9.94&   0.50& 00&    0.0 \\
 11024&   24.4&   10&   7.87&   6.72&   6.36&   0.95& B&  10.29&   81&  10.39&   8.31&   7.61&   0.70& 11&    2.0 \\
 11324&   19.0&    6&   9.26&   8.15&   7.88&   0.75& B&  15.33&  165&   9.37&   8.39&   8.12&   0.71& 10&    2.0 \\
\enddata                                                                                                                                  
~  \\  
\end{deluxetable}                                                                                                                         

\begin{deluxetable}{ll} 
\tabletypesize{\scriptsize}   
\tablecaption{Notes on individual targets
\label{tab:notes} }
\tablewidth{0pt}
\tablehead{ \colhead{HIP} & \colhead{Text} }
\startdata  
  9499&   A is SB, no orbit. N04: $\Delta RV$=70.7. The B-companion is optical.  \\
 10305&   A is SB2, period 94.788d, resolved by speckle (TOK 39). A is evolved. B = HIP 10303 is physical.  \\
 10579&   AB = HJ 3491 is physical. Also CPM with HIP 10754 at 836$''$. \\
 10621&   NOMAD: PM(B)=(0,$-$30), PM(A)=(18,60),  AB is optical. BC is a binary at 2\farcs54, 180.0$^\circ$, $\Delta K$=0.47, $\Delta V$=0.89. \\
      &   There is a CPM companion HIP 10754 at 836$''$.  \\
 10710&   No data. The PSC companion at 14\farcs4 is likely physical based on the CMD and $N^*$=13.  \\
 11324&   B = HIP 11323, physical. Both components are below the MS, the true parallax should be 12.6mas \\
      & (as for the B-component in Hipparcos). \\
\enddata                                                                                                                                  
\end{deluxetable}

\begin{deluxetable}{ccc ccc} 
\tabletypesize{\scriptsize}   
\tablecaption{Companion count
\label{tab:count} }
\tablewidth{0pt}
\tablehead{ $J_{\rm abs}$ & $n_p$ & $n_o$ & $n_c$ & $\alpha$ & $q_0$  }
\startdata  
  6--7 & 61 & 20  & $51 \pm 8$  &  0.84 & 0.51 \\
  7--8 & 45 & 27  & $31 \pm 7$  &  0.70 & 0.34 \\ 
  8--9 & 41 & 51  & $15 \pm 7$  &  0.38 & 0.19 \\
 9--10 & 59 & 107 & $ 5 \pm 9$  &  0.09 & 0.12 \\
\enddata                                                                                                                                  
\end{deluxetable}

\end{document}